\documentclass[11pt,a4paper]{article}
\usepackage{jcappub}

\usepackage{bm}
\usepackage{epsfig}
\usepackage{citesort}
\usepackage{graphicx}
\usepackage{amsmath}
\usepackage{amssymb}
\usepackage{amsbsy}
\usepackage{color}
\usepackage{subfigure}
\usepackage{slashed}
\usepackage{afterpage}
\usepackage{psfrag}
\usepackage{axodraw}
\usepackage{hyperref}
\renewcommand\vec[1]{{\bf #1}}
\newcommand\vecs[1]{\ensuremath\boldsymbol{#1}}

\renewcommand\({\left(}
\renewcommand\){\right)}

\newcommand{\be}{\begin{equation}}
\newcommand{\ee}{\end{equation}}
\newcommand{\bea}{\begin{eqnarray}}
\newcommand{\eea}{\end{eqnarray}}
\def\nn{\nonumber}

\renewcommand\({\left(}
\renewcommand\){\right)}

\newcommand{\bra}[1]{\langle #1 |}
\newcommand{\ket}[1]{| #1 \rangle}

\newcommand{\exclude}[1]{}

\definecolor{gre}{rgb}{0,0.4,0.3}

\begin{document}
\subheader{\hfill MPP-2011-219}

\title{Solar axion flux from the axion-electron coupling}

\author[a,b]{Javier~Redondo}

\affiliation[a]{Arnold Sommerfeld Center, 
Ludwig-Maximilians-Universit\"at, Theresienstr.~37, D-80333 M\"unchen, Germany}
\affiliation[b]{Max-Planck-Institut f\"ur Physik, %(Werner-Heisenberg-Institut)
F\"ohringer Ring 6, D-80805 M\"unchen, Germany}

%\emailAdd{paola.arias@desy.de}
%\emailAdd{cadamuro@mppmu.mpg.de}
%\emailAdd{mark.goodsell@desy.de}
%\emailAdd{joerg.jaeckel@durham.ac.uk}
%\emailAdd{andreas.ringwald@desy.de}
\emailAdd{redondo@mpp.mpg.de}

\abstract{In non-hadronic axion models, where axions couple to electrons at tree level, the solar axion flux is completely dominated by the ABC reactions (Atomic recombination and deexcitation, Bremsstrahlung and Compton). 
In this paper the ABC flux is computed from available libraries of monochromatic photon radiative opacities (OP, LEDCOP and OPAS) by exploiting the relations between axion and photon emission cross sections. 
These results turn to be $\sim 30\%$ larger than previous estimates due to atomic recombination (free-bound electron transitions) and deexcitation (bound-bound), which where not previously taken into account. 
}
 
\maketitle

\section{\label{intro}Introduction and summary of results}

The axion is a hypothetical $0^-$ boson predicted by the Peccei-Quinn solution of the strong CP problem of the standard model~\cite{Peccei:1977hh,Peccei:2006as,Weinberg:1977ma,Wilczek:1977pj}, see~\cite{Kim:2008hd} for a review. 
It arises as the pseudo-Nambu-Goldstone boson of a global symmetry, U$_{\rm PQ}$(1), which is spontaneously broken at high energy scale $f_a$ and only violated at the quantum level by the colour anomaly~\cite{Kim:1979if,Shifman:1979if,Dine:1981rt,Zhitnitsky:1980tq,Pi:1984pv}.
All values of $f_a$  solve the strong CP problem but the axion couples to hadrons, photons and leptons with interaction strengths inversely proportional to $f_a$ and thus the smallest values of $f_a\lesssim 10^9$ GeV have been experimentally excluded by direct laboratory searches~\cite{Donnelly:1978ty,Barshay:1981ky,Barroso:1981ta,Peccei:1981za,Krauss:1986wx,Bardeen:1986yb,Riordan:1987aw}, constraints on hot dark matter~\cite{Archidiacono:2013cha,Cadamuro:2010cz,Hannestad:2010yi,Hannestad:2008js,Hannestad:2007dd} and stellar evolution~\cite{Raffelt:1996wa,Gondolo:2008dd,Friedland:2012hj,Cadamuro:2011fd,Raffelt:1994ry,RaffeltViaux,Raffelt:1985nj,Isern:1992,Isern:2010wz,Isern:2008nt,Melendez,Raffelt:1987yt,Turner:1987by,Mayle:1987as,Mayle:1989yx,Umeda:1997da,Keller:2012yr}. 
Axions get a calculable mass by mixing with the pseudoscalar mesons, $m_a/$meV$\simeq 6\times 10^9{\rm GeV}/f_a$, so they are extremely weakly interacting, extremely light and connected to very high energy physics. 
They are cosmologically stable and can be copiously produced in the early universe by non-thermal mechanisms, becoming splendid candidates for the cold dark matter of the universe~\cite{Preskill:1982cy,Abbott:1982af,Sikivie:1982qv,Hiramatsu:2010yn,Wantz:2009it}. 
There are good prospects for detecting dark matter axions in the $m_a\sim 2-20\ \mu$eV range~\cite{Asztalos:2009yp,ADMXprospects,ADMXprospects2} with resonant microwave cavities (the haloscope experiments of Sikivie~\cite{Sikivie:1983ip}). 
New experimental techniques, recently proposed, can extend this window~\cite{Graham:2011qk,Graham:2013gfa,Budker:2013hfa,Horns:2012jf,Jaeckel:2013sqa,Jaeckel:2013eha}.

Beyond the dark matter window, finding the axion experimentally seems only plausible in the meV mass frontier~\cite{Raffelt:2011ft}  with a next generation axion helioscope~\cite{Irastorza:2011gs}. 
A helioscope~\cite{Sikivie:1983ip} aims at detecting the flux of axions emitted from the Sun, the brightest source of axions in the sky. 
It consists in a pipe pointing the Sun in which a very intense transverse magnetic field acts 
as as axion-photon mixing agent, triggering the conversion of solar axions into detectable X-rays~\cite{vanBibber:1988ge}. 
Three such experiments have been performed~\cite{Lazarus:1992ry,Inoue:2002qy,Moriyama:1998kd,Inoue:2008zp}, the still-ongoing CERN axion solar telescope (CAST) still holding the most restrictive experimental constraints on the axion-photon coupling~\cite{Zioutas:2004hi,Andriamonje:2007ew,Arik:2008mq,Arik:2011rx,Arik:2013nya}. 
The experience gained with CAST has led the conception of an scaled-up version that can be sensitive to 
meV-mass axions. A community is forming around this future experiment, named the International AXion Observatory (IAXO)~\cite{Irastorza:2012qf}, and the first steps towards a technical design have been taken~\cite{Shilon}. 

Axions with masses in the multi-meV mass range can play a noticeable role in stellar evolution, in particular in the cooling of compact objects such as red-giant cores~\cite{Raffelt:1994ry,RaffeltViaux}, white dwarfs~\cite{Raffelt:1985nj,Isern:1992,Isern:2010wz,Isern:2008nt}, supernova cores~\cite{Raffelt:1987yt,Turner:1987by,Mayle:1987as,Mayle:1989yx} and neutron stars~\cite{Umeda:1997da,Keller:2012yr}. In fact, the most restrictive limits on the axion couplings to nucleons, photons and  electrons come from the reasonable agreement of astronomical observations with standard stellar-cooling mechanisms: photon surface cooling and neutrino emission from dense cores. Axion emission can speed up enormously stellar cooling and spoil badly this agreement --- hence the strong and robust bounds --- but it can also be used to reduce slight discrepancies between observations and predictions. 
Such are the cases for white dwarfs~\cite{Isern:2008nt,Corsico:2012ki,Corsico:2012sh} and red-giant stars in the globular cluster M5~\cite{RaffeltViaux}, where small discrepancies can be mitigated by introducing axions with a Yukawa coupling to electrons $g_{ae}\sim 10^{-13}$, a natural value for meV-mass axions. 
Let us recall that in all mentioned cases, the preference for anomalous cooling is statistically not very significant  and might be due to unaccounted systematics or neglected standard effects. Clearly, the situation will benefit from direct experimental verification and here, the Sun and IAXO might be our best allies.

\begin{figure}[htbp]
\begin{center}
\includegraphics[width=12cm]{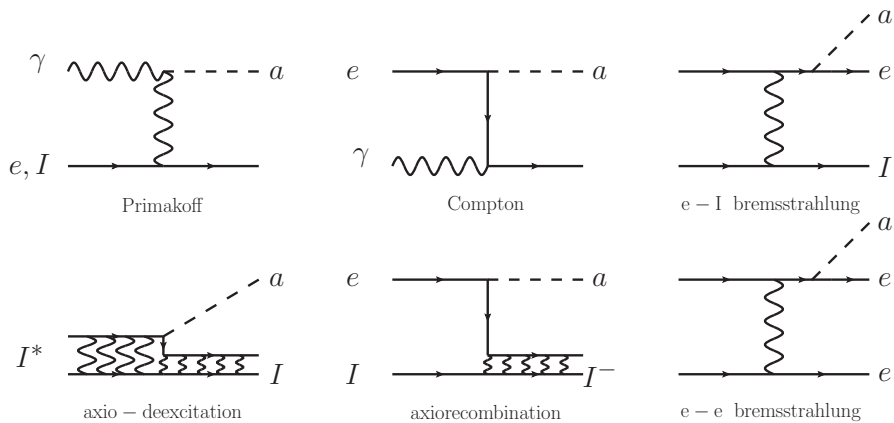}
\caption{ABC reactions responsible for the solar axion flux in non-hadronic axion models. }%\small Reactions responsible for the }
\label{fig:reactions}
\end{center}
\end{figure}

A prime theoretical input for helioscopes is the solar axion flux. 
The solar interior is a well-understood weakly coupled plasma which permits relatively precise calculations of 
axion production reactions. The most important parameters that determine the axion flux are the axion-two-photon coupling and the axion-electron coupling. The first drives the Primakoff production of axions in photon collisions with charged particles of the solar plasma, $\gamma+q\to a+q$, and has been thoroughly studied~\cite{Dicus:1978fp,Raffelt:1985nk,Raffelt:1987np}. The Primakoff flux is dominant in hadronic axion models such as the KSVZ~\cite{Kim:1979if,Shifman:1979if} where the axion-electron coupling is absent at tree level. 
In generic models, the axion-electron coupling can appear at tree level, and in grand unified theories (GUTs) is unavoidable. The axion-electron coupling drives a number of reactions of comparable importance that completely overshadow the Primakoff flux in non-hadronic axion models. 
The most important are the ABC reactions: 
{\bf A}tomic axio-recombination~\cite{Dimopoulos:1986kc,Dimopoulos:1986mi,Pospelov:2008jk} and {\bf A}tomic axio-deexcitation, 
axio-{\bf B}remsstrahlung in electron-Ion~\cite{Zhitnitsky:1979cn,Raffelt:1985nk,Krauss:1984gm} or electron-electron collisions~\cite{Raffelt:1985nk}, 
{\bf C}ompton scattering~\cite{Mikaelian:1978jg,Fukugita:1982ep,Fukugita:1982gn}, 
see figure~\ref{fig:reactions} for a sample of Feynman diagrams. 

The axion flux from ABC processes has received less attention than the Primakoff. 
After its identification by Krauss, Moody and Wilczek~\cite{Krauss:1984gm} it became clear that 
electron-Ion (mostly hydrogen and helium) and electron-electron axio-bremsstrahlung dominate the solar flux from the axion-electron coupling~\cite{Raffelt:1985nk}. 
Axio-recombination and atomic de-excitation are significant only for ions of metals\footnote{In the context of stellar evolution everything but hydrogen and helium is a metal.} which are much less abundant than hydrogen, helium or electrons. 
In 1986, Dimopoulos, Starkman and Lynn, pointed out that the axio-recombination cross sections are actually much larger than those of bremsstrahlung and Compton (for electron energies typical of the solar interior) and thus they partially compensate the low metal content of the Sun~\cite{Dimopoulos:1986kc}. 
Still, their first estimate found the axio-recombination flux to be largely subdominant~\cite{Dimopoulos:1986kc}. 
However, that calculation included  K-shell recombination only and employed cross sections with excessive screening which were additionally underestimated by a factor 1/2 recently pointed by Pospelov, Ritz and Voloshin~\cite{Pospelov:2008jk} (see also~\cite{Derevianko:2010kz}). 
All these facts and the omission of atomic axio-deexcitation, suggests that actually the role of metals in the solar axion flux might be larger than previously calculated and motivates to undertake a new estimation of axio-recombination and axio-deexcitation. 
Taken at face value, this is a gargantuan task which involves understanding the energy levels and occupation probabilities of a large number of atomic states for each nucleus inside the Sun and then computing cross sections and transition probabilities. 

The main point of this paper is to show that there is no need to face a direct calculation because all the ingredients are already available: 
\begin{itemize}
\item The spin-averaged differential cross section of emitting an axion in an atomic transition $|e_i\rangle\to |e_f\rangle$ 
is proportional to the analogous cross section for emitting a photon in the same transition. 
Thus, axion and photon production rates as function of energy, $\omega$, are proportional in the solar plasma.  
\item In a plasma in thermal equilibrium, photon absorption and production rates are related by detailed balance.
\item Photon absorption coefficients (radiative opacities) have been the subject of extensive research efforts because of their central role in stellar evolution and plasma diagnosis. 
Libraries of monochromatic opacities for different nuclei are available in wide ranges of temperature and density.  
We can use therefore their state-of-the-art calculations to get the photon absorption rates for any point inside the Sun. 
\end{itemize}

\begin{figure}[htbp]
\begin{center}
\includegraphics[width=9cm]{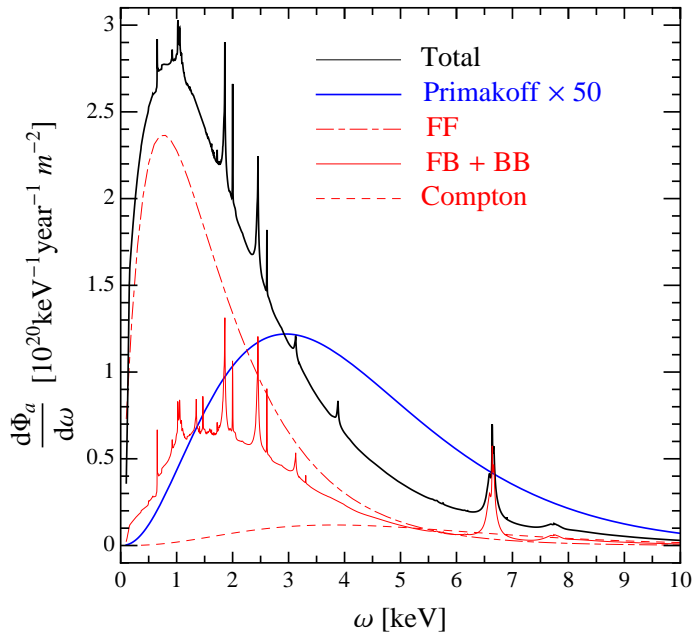}
\caption{Flux of solar axions due to ABC reactions driven by the axion-electron coupling (for $g_{ae}=10^{-13}$). 
The different contributions are shown as red lines: Atomic recombination and deexcitation (FB+BB, solid), Bremsstrahlung (FF, dot-dashed) and Compton (dashed). 
The Primakoff flux from the axion-photon coupling is shown for comparison using $g_{a\gamma}=10^{-12}$, a typical value for meV axions having  $g_{ae}=10^{-13}$. Note that has been scaled up by a factor 50 to make it visible. }%\small Reactions responsible for the }
\label{fig:flux}
\end{center}
\end{figure}

In this paper we put together these three ingredients to provide a new and more complete estimate of the full ABC flux of solar axions. The result is shown in figure~\ref{fig:flux}, where we also show the A, B and C contributions separately. 
%The relative importances of ei,ee , BF+BB and Compton are for the number flux: $55.2::9.7::27.6::7.6$ and the energy flux $44::7.4::33::15.6$. 
The relative importances for the total fluxes $\Phi$ and the luminosities $L$ (energy flux) are 
\bea
\Phi_B:\Phi_C:\Phi_A &=& 64.9:7.6:27.6 \\
L_B:L_C:L_A &=& 51.4:15.6:33 	\,  . 
\eea
The A-processes contribute $\sim 27.6\%$ to the total flux ($33\%$ for the luminosity), raising previous estimates by a factor of $\sim 1.3$ ($1.5$ for the luminosity). 
These are not dramatic changes but they are certainly non negligible. 

The results of this paper will allow a more precise search for solar axions. 
The ABC flux peaks at $\omega\sim$ keV and this is where IAXO sensitivity will be larger. 
Here the new estimate for the flux is $\sim 20\%$ larger than used in~\cite{Irastorza:2011gs} so the IAXO sensitivity to $g_{ae}$ is expected to increase by $10\%$. 
Note that the CAST collaboration has already used these results to constrain the product of the axion-electron coupling (responsible for axion production) and axion-photon coupling (responsible for detection in CAST and the future IAXO) $g_{ae}\times g_{a\gamma}< 8.1\times10^{-23}\,\rm GeV^{-1}$~\cite{Barth:2013sma}. 

Axio-recombination and axio-deexcitation can in principle be also significant in the cooling of white dwarves and red giants. 
The method proposed in this paper allows to easily compute these fluxes for any stellar plasma for which radiative opacities are available. 
However, in degenerate plasmas screening is much stronger than in the Sun's core and most of the bound states can be effectively screened to the continuum. Thus we shall not expect big changes. 

For completeness, let us remind that the solar axion luminosity is constrained to be smaller than $10\%$ of the solar luminosity $L_\odot$~\cite{Gondolo:2008dd}, for larger values require faster nuclear reactions accompanied by a large flux of Boron neutrinos which is excluded by SNO. In the original paper~\cite{Gondolo:2008dd}, the authors included only bremsstrahlung and Compton and thus the axion luminosity was underestimated by a factor of 2/3. Correcting for this factor, the axion-electron coupling is now constrained to be
\be
g_{ae} < 2.3\times 10^{-11} \, .
\ee
However, the constrain is superseded by the white dwarf and red giant arguments and thus our improvement is largely irrelevant. 

After having presented the novel ideas and results, we devote the rest of this paper to details. 
In section 2 we develop the calculation framework and discuss details and subtleties one has to take into account. 
In section 3 we compute the solar axion flux with different opacity codes, OP, LEDCOP and OPAS and find very good agreement. Finally, in section 4 we discuss possible applications for hypothetical particles other than axions.

%%%%%%%%%%%%%%%%%%%%%%%%%%%%%%%%%%%%%%%%%%%%%%
\section{ABC processes from photon opacity}\label{ABC}
%%%%%%%%%%%%%%%%%%%%%%%%%%%%%%%%%%%%%%%%%%%%%%

In this section we develop in detail the machinery that allows to use monochromatic opacities to compute the axion flux. The interaction lagrangian densities for axions and photons with electrons are
\bea
{\mathcal L}_a 		&=& 	g_{ae} \frac{\partial_\mu a}{2 m_e} \bar \psi_e \gamma^\mu\gamma_5 \psi_e	\equiv  
- i g_{ae} a \bar \psi_e \gamma_5 \psi_e\\ 
{\mathcal L}_\gamma 	&=&  	e A_\mu \bar \psi_e \gamma^\mu \psi_e 									. 
\eea
where $a$ is the axion field, $A_\mu$ the electromagnetic vector potential, $\psi_e$ the electron spinor field, $m_e$ the electron mass, $e\sim 0.3$ the electron charge (the fine structure constant is $\alpha=e^2/4\pi$) and $g_{ae}$ the axion-electron Yukawa coupling.  
The latter is given by 
\be
\label{gae}
g_{ae} = C_{ae} \frac{m_e}{f_a}
\ee
where $f_a$ is the axion decay constant and $C_{ae}\sim {\cal O}(1)$ a model-dependent coefficient. 
We work in Lorentz-Heaviside units and take $\hbar=c=k_\text{B}=1$.

\subsection{Axion emission}

We define the production rate such that the number of axions created per unit time and volume 
is given by the phase-space integral of $\Gamma_a^{\rm P}$
\be
\frac{d N_a}{dV dt } = \int \frac{d^3 \mathbf{k}}{(2\pi)^3} \Gamma_a^{\rm P}(\omega)=
\int_0^\infty \frac{\omega^2 d \omega}{2 \pi^2} \Gamma_a^{\rm P}(\omega).
\ee

The emission of axions happens mainly from the processes depicted in figure~\ref{fig:reactions}  
\be
\Gamma_a^{\rm P}(\omega) = 
\Gamma^{\rm ff}_a+
\Gamma^{\rm fb}_a+
\Gamma^{\rm bb}_a+
\Gamma^{\rm C}_a +
\Gamma^{\rm ee}_a , 
\ee
where the different contributions are 
\begin{itemize}
\item[(ff)] electron-ion bremsstralung, also known as free-free electron transitions
\be \nn e+I\to e+I+a , \ee  
\item[(bf)] atomic axio-recombination, also known as electron capture or free-bound electron transitions
\be \nn e+I\to I^-+a , \ee 
\item[(bb)] atomic deexcitation, or bound-bound electron transitions
\be \nn I^*\to I+a , \ee
\item[(C)] Compton-like scattering 				(photo-production) 
\be \nn e+\gamma \to e+a  , \ee
\item[(ee)] electron-electron bremsstrahlung
\be \nn e+e\to e+e+a .\ee
\end{itemize}

\subsection{Photon opacity}

In the language of radiative transfer\footnote{The physics of radiative transfer is covered in many standard text books~\cite{Chandrasekhar1939,Chandrasekhar1950,Mihalas1978,Rybicki1979}. The lecture notes~\cite{Rutten2003} are also very accessible.}, the specific absorption coefficient $k(\omega)$ is defined through the energy transport equation, which rules the damping and sourcing of a pencil of radiation of specific intensity $I_\omega$ along a line of sight within a plasma
\be
\frac{d I_\omega}{d s} = - k(\omega) I_\omega + j(\omega)
\ee
where $j(\omega)$ is the source. 
The radiative opacity $\kappa(\omega)$ is defined as the absorption coefficient per unit mass of target  $\kappa(\omega)=k(\omega)/\rho$ with $\rho$ the density of the medium.  

In the solar interior, the absorption coefficient has contributions from true absorption processes (free-free, bound-free and bound-bound electronic atomic transitions) and Compton scattering~ 
\be
k(\omega) = \(\Gamma_\gamma^{\rm A,ff}+
\Gamma_\gamma^{\rm A,bf}+
\Gamma_\gamma^{\rm A,bb}\) (1-e^{-\omega/T})+\Gamma^{\rm A,C}_\gamma . 
\ee
Note that the factor $(1-e^{-\omega/T})$ corrects for stimulated emission in a bath in thermal equilibrium and appears 
only in the true absorption processes. 
The true absorption processes correspond one-to-one to the first three (time-reversed) axion emission processes listed above by substituting the axion for a photon. 

The value of $\Gamma^{\rm A}_\gamma$ depends upon the densities of the different atomic species $n_Z$ and temperature $T$ and requires a huge numerical machinery. 
For instance the bound-free contribution can be calculated as
\be
\Gamma_\gamma^{\rm A,bf}=\sum_Z n_Z \sum_s r_s \sigma (\gamma+Z_s\to Z_{s'}+e^{-}) 
\ee
where $n_Z$ are the densities of atoms of nuclear charge $Z$, $r_s$ the fraction of these atoms in the state $s$ and 
$\sigma (\gamma+Z_s\to Z_{s'}+e^{-}) $ the total photo-ionization cross section. The calculation of $r_s$ involves solving the atomic structure in a relatively-dense medium, solving the Saha equation for the ionization fraction and computing the partition functions for the probability of initial states. The cross section has to include the non-trivial atomic structure, electrostatic screening, Coulomb wavefunctions in the final states when applicable and so forth.

Fortunately, photon opacities are routinely calculated and improved for their use in stellar evolution and 
general plasma physics. Many opacity databases are publicly available although most of them provide only the frequency-averaged opacities relevant for radiative transfer in stellar structure. 
Only a few provide monochromatic opacities, among which we can highlight the Opacity Project (OP)~\cite{OP} and the   ``Los Alamos Light Element Detailed Configuration OPacity code''~(LEDCOP)~\cite{LEDCOP}. We will describe some of their properties and assumptions in the next section. 

In order to use photon opacities, we have to relate them to photon production rates. 
In thermal equilibrium, the rate of a reaction and its inverse are related by detailed balance 
\be
\Gamma^{{\rm A},p}_\gamma(\omega)=e^{\omega/T}\Gamma^{{\rm P},p}_\gamma(\omega) \, .   
\ee
Note that we define $\Gamma^{{\rm P},i}_\gamma(\omega)$ as the production rate of photons 
per phase-space volume, averaged over polarizations. The total production rate is $2\times \Gamma^{{\rm P},i}_\gamma(\omega)$, accounting for the two photon polarizations. 

A crucial point for this paper is that this equality holds \emph{for each process separately}. 
 
\subsection{Relations between photon and axion emission processes}

We can now exploit the relation between the emission rates of photons and axions 
to express $\Gamma^{\rm P}_a$'s as a function of $\Gamma^{\rm P}_\gamma$'s. 
These relations depend on the process under consideration and have to be considered case by case. 
Fortunately, we need only to consider three different cases.  

%Case I applies when axion/photon production is triggered by an electron undergoing an atomic transition. The axion and photon production rates are proportional, and the proportionality factor is the same for all atomic transitions relevant in the Sun: ff, bf and bb. 
%Axion and photon emission in electron-electron bremsstrahlung are not subject to this proportionality and have to be discussed aside (Case II). Finally, axion/photon production rates in Compton scattering are also proportional, but the proportionality constant is a factor of 2 stronger than case I deserving a separate case III.   

%%%%%%%%%%%%%%%%%%%%%%%%%%%%%%%%%%%%%%%%%%%%%%%%%%%%%%
\subsubsection*{Case I: ff,fb,bb processes}
%%%%%%%%%%%%%%%%%%%%%%%%%%%%%%%%%%%%%%%%%%%%%%%%%%%%%%

In processes in which a photon/axion is emitted when an electron makes an atomic transition $e_i\to e_f$, i.e. in the interaction of one electron with an ion, the spin-averaged matrix element of emitting an axion of energy $\omega$ is proportional to the analogous matrix element of emitting a photon (also polarization averaged) of the same energy, 
\be
\label{eq:matrixequality}
\frac{\sum_{s_i,s_f}|{\cal M}(e_i\to e_f+a)|^2}{\frac{1}{2}\sum_{\epsilon} \sum_{s_i,s_f}|{\cal M}(e_i\to e_f+\gamma)|^2} = \frac{1}{2}\frac{g_{ae}^2 }{e^2 }\frac{\omega^2}{m_e^2}
%=\frac{ g_{ae}^2 \omega^2}{16\pi \alpha m_e^2}%=\frac{C_{ae}^2}{16\pi \alpha}\(\frac{\omega}{f_a}\)^2 
.
\ee

The derivation of~\eqref{eq:matrixequality} was first presented in~\cite{Dimopoulos:1986mi} and later corrected in~\cite{Pospelov:2008jk,Derevianko:2010kz}, see also the appendix of~\cite{Derevianko:2010kz}.  
This equality is based on three basic approximations:   
1) non-relativistic expansion of the interaction hamiltonians, 2) initial and final states are separable in spatial and spin wave-functions and 3) multipole expansion of the transition amplitudes. % $e^{i \vec k\cdot\vec X}= 1+i \vec k\cdot\vec X+...$. 
The photon emission is well described by the electric dipole approximation ($e^{i \vec k\cdot\vec X}\simeq 1$) 
\bea \nn
\langle f|H_{I,}^{\gamma}|i\rangle &\sim& - 2  e 
\bra{f}e^{i \vec k\cdot\vec X}\(
 \vecs \epsilon \cdot  \vec P + i {\vec S}\cdot(\vec k\times \vecs \epsilon) \) \ket{i}	 \sim	\\
 &&  - 2  e \bra{f}\( \vecs \epsilon \cdot  \vec P \) \ket{i}
    = - 2 i e m_e \omega   \bra{f}\( \vecs \epsilon \cdot  \vec X \) \ket{i}, 
\eea 
($\mathbf{X},\mathbf{P},\mathbf{S}$ are the electron's position, momentum and spin operators) 
while for axion emission one needs to retain one more order 
\bea \nn
\langle f|H_{I}^{a}|i\rangle &\sim& - 2 g_{ae}  
\bra{f}e^{i \vec k\cdot\vec X}\(
 \vec k \cdot \vec S - \frac{\omega}{m_e}\vec P \cdot \vec S\) \ket{i}	
 \sim  														\\
 && - 2 g_{ae} \omega^2 i   
\bra{f}\( (\vec n\cdot\vec X) (\vec n \cdot \vec S) - \vec X \cdot \vec S\) \ket{i}		, 
\eea
where $\bf n$ is a unit vector in the direction of the axion momentum $\bf k$. 
Note that, as a further simplification eq. ~\eqref{eq:matrixequality} assumes that photons and axions can both be taken to be massless. The generalization for non-zero axion masses does not present difficulties, see~\cite{Dimopoulos:1986mi,Pospelov:2008jk,Derevianko:2010kz}.

Integrating over the phase space of final states, we find that the same proportionality factor of the matrix elements~\eqref{eq:matrixequality} applies to the emission cross sections (the latter averaged over photon-polarizations)
\be
\label{eq:crosssection}
\frac{\sigma(e_i+I\to e_f+I+a)}{\frac{1}{2}\sigma(e_i+I\to e_f+I+\gamma)} = \frac{1}{2}\frac{g_{ae}^2}{e^2 }\frac{\omega^2}{m_e^2}
%= \frac{ g_{ae}^2 \omega^2}{16\pi \alpha m_e^2}%=\frac{C_{ae}^2}{16\pi \alpha}\(\frac{\omega}{f_a}\)^2 
.
\ee
and, after convolving these with the appropriate densities of initial states, also for the  
thermal axion/photon emission rates
\be
\label{eq:gammas}
\frac{\Gamma^{{\rm P},i}_a(\omega)}{\Gamma^{{\rm P},i}_\gamma(\omega)} = \frac{1}{2}\frac{g_{ae}^2 \omega^2}{e^2 m_e^2}%= \frac{ g_{ae}^2 \omega^2}{16\pi \alpha m_e^2}
\quad ;\quad i=\rm ff,\ fb,\ bb
\ee
because kinematics are the same for axions and photons in the massless limit. 

The initial and final states can be bound or unbound, i.e. this formula applies for axion/photon emission in free-free, free-bound or bound-bound transitions for which the above approximations are good. 
The reason is that in reactions in which the electrons are non-relativistic, the electron momenta involved ($p$) are typically larger than the emitted axion/photon energy. 
In a free-free transition, energy conservation implies that $\omega \simeq(p_f^2-p_i^2)/2 m_e$, which is much smaller than $p_f-p_i$ because $(p_i+p_f)/2m_e$ is small. 
For free-bound or bound-bound transitions, the electron momenta involved are of order of the inverse orbital sizes $p_n\sim Z\alpha m_e/n$ ($Z$, ion charge, $n$ principal number) while photon energies are similar to atomic energies $E_n\simeq Z^2\alpha^2/2n^2$ (unless the initial electron energy is much larger than $E_n$ in which case $\omega \sim p_i^2/2m_2$ and again smaller than $p_i$). 
In either case, it seems reasonable to neglect $\omega$ when terms of order $p$ are present, and consider that factors $\vec k\cdot \vec x$ are $\ll1$ when convoluted with atomic wavefunctions.    

The discrepancy of a factor 2 found in~\cite{Pospelov:2008jk} has its origin in the electric dipole term in the axion emission calculation, $\propto - \omega \vec P \cdot \vec S/m_e$. This was neglected in~\cite{Dimopoulos:1986mi,Dimopoulos:1986kc} because it is zero in the strict non-relativistic limit\footnote{This has propagated to further literature, particularly to standard text books~(note eq. (3.7) in~\cite{Raffelt:1996wa}) }. However, it contributes as much as $\vec k \cdot \vec S$, changing the relation \eqref{eq:matrixequality}. Let us emphasise that this formula is not only valid for axion/photon emission in free-bound transitions but also for free-free and bound-bound. 

Finally, we can explicitly cross-check this formula against the calculations present in the literature for the case of free-free transitions. 
The photon production rate in electron collisions with ionized nuclei of electric charge $Z e$ and number density $n_Z$, including Debye screening in the Born approximation is (see e.g. eq. (4.5) of~\cite{Redondo:2013lna}) 
\be
\label{eq:axioBremsstrahlung}
\Gamma^{\rm P,ff}_\gamma(\omega)= \alpha^3 Z^2 \frac{64\pi^2  }{3\sqrt{2\pi} } \frac{n_Z n_e}{\sqrt{T} m_e^{3/2}\omega^3}e^{-\omega/T}  F(w,y), 
\ee
where 
\be
F(w,y) = \int_0^\infty dx\, x\, e^{-x^2} \int_{\sqrt{x^2+w}-x}^{\sqrt{x^2+w}+x}\frac{t^3 dt}{(t^2+y^2)^2} , 
\ee
with $y=k_{\rm s}/\sqrt{2 m_e T}$ where $k_s$ the Debye screening scale. 
The axion production rate with the same assumptions was computed in~\cite{Raffelt:1985nk}. 
The formula for the emission rate per unit volume can be found in~\cite{Raffelt:PhD} or translated from eq. (40) of~\cite{Raffelt:1985nk}. In our notation it is
\be
\Gamma^{\rm P,ff}_a(\omega)= \alpha^2 g_{ae}^2 Z^2 \frac{8 \pi }{3\sqrt{2\pi } } \frac{n_Z n_e}{\sqrt{T}m_e^{7/2
}\omega}e^{-\omega/T}  F(w,y)
\ee
Dividing both expressions (and using $\alpha=e^2/4\pi$) we obtain~\eqref{eq:gammas}. 

%%%%%%%%%%%%%%%%%%%%%%%%%%%%%%%%%%%%%%%%%%%%%%%%%%%%%%
\subsubsection*{Case II: e-e bremsstrahlung}
%%%%%%%%%%%%%%%%%%%%%%%%%%%%%%%%%%%%%%%%%%%%%%%%%%%%%%
The cross-section of photon bremsstrahlung in a e-e collision is zero in the electric-dipole approximation, simply because the electric dipole of two colliding electrons in the centre of mass is zero. Photon emission happens at the quadrupole level which is much suppressed with respect to e-I processes~\cite{Maxon:1967}. For this reason $\Gamma^{\rm A,ee}_\gamma$ is often neglected for the calculation of opacities at conditions of the solar interior, or included as a $\mathcal{O}(10^{-3})$ correction to $\Gamma^{\rm A,ff}_\gamma$~\cite{Iglesias:1996}. 
However, axion emission in e-e collisions is of the same order than in e-I collisions~\cite{Raffelt:1985nk} and thus has to be included by hand. 
The emission rate was computed by Raffelt~\cite{Raffelt:PhD,Raffelt:1985nk} 
\be
\label{eq:axioBremsstrahlungee}
\Gamma^{\rm P,ee}_a=\alpha^2 g_{ae}^2 \frac{4 \sqrt{\pi} }{3} \frac{n^2_e}{\sqrt{ T} m_e^{7/2
}\omega}e^{-\omega/T}  F(w,\sqrt{2}y) . 
\ee  

The simple proportionality \eqref{eq:gammas} fails in this case,  
which shall not surprise anyone because the approximations required for its derivation do not apply. 
In particular, here we have two electrons radiating photons while \eqref{eq:gammas} applies for only one. 
Each electron is accelerated and thus indeed radiates dipole radiation, however the emission of the two electrons has the opposite phase and cancels. %This effect cannot be taken into account in  \eqref{eq:gammas} in a simple way.  

\subsection*{Case III: Compton scattering }
 In the non-relativistic limit, the cross section of photo-production of axions in Compton-like scattering is 
 $\sigma_{C,a}=\alpha g_{ae}^2\omega^2/3 m_e^4$~\cite{Raffelt:1996wa,Mikaelian:1978jg}. 
Explicitly the production rate is 
\be
\label{eq:axioCompton}
\Gamma^{{\rm P},C}_a(\omega) =  \frac{\alpha g_{ae}^2 \omega^2}{3 m_e^2}\frac{n_e }{e^{\omega/T}-1} . 
\ee
The cross section of Compton scattering in the non-relativistic limit is the Thomson cross section $\sigma_{C,\gamma}=8\pi \alpha^2/3 m_e^2$, 
we find for the ratio have
\be
\frac{\Gamma^{{\rm P},C}_a(\omega)}{\Gamma^{{\rm P},C}_\gamma(\omega)} = \frac{f(\omega) n_e \sigma_{C,a}}{f(\omega) n_e \frac{1}{2}\sigma_{C,\gamma}}=\frac{g_{ae}^2 \omega^2}{e^2 m_e^2} .
\ee
Note the factor of $2$ difference between this and \eqref{eq:gammas}, which makes Compton relatively more
efficient than ff, fb or bb processes for emitting axions (relative to the photon emission).  

The Feynman diagrams for Thomson scattering are somewhat analogous to those of bremsstrahlung and one might ask why \eqref{eq:gammas} does not apply. The reason is that in Thomson scattering 
the electron experiences a very small momentum transfer of the order of the photon energy radiated 
$|\Delta \mathbf{p}|\approx \omega$ and \eqref{eq:gammas} requires 
$|\Delta \mathbf{p}|\gg \omega$, which is guaranteed in the typical ff, fb and bb transitions we consider for the solar opacity and the ABC axion flux.  

\subsection*{Useful formulas}

Putting together all the pieces, the axion production rate in ABC processes can be expressed in terms of the photon absorption coefficient as
\bea \nonumber
\Gamma^{\rm P}_a(\omega) &=& \frac{1}{2}\frac{g_{ae}^2 \omega^2}{e^2 m_e^2}\(\Gamma^{\rm P,ff}_\gamma +\Gamma^{\rm P,bf}_\gamma+\Gamma^{\rm P,bb}_\gamma\)+\Gamma^{\rm P,C}_a+\Gamma^{\rm P,ee}_a \\ 
&=& \nonumber
\frac{1}{2}\frac{g_{ae}^2 \omega^2}{e^2 m_e^2}\(\Gamma^{\rm A,ff}_\gamma +\Gamma^{\rm A,bf}_\gamma+\Gamma^{\rm A,bb}_\gamma\)e^{-\omega/T}+\Gamma^{\rm P,C}_a+\Gamma^{\rm P,ee}_a \\ 
&=& \nonumber
\frac{1}{2}\frac{g_{ae}^2 \omega^2}{e^2 m_e^2}\(\Gamma^{\rm A,ff}_\gamma +\Gamma^{\rm A,bf}_\gamma+\Gamma^{\rm A,bb}_\gamma+\frac{\Gamma^{\rm A,C}_\gamma}{1-e^{-\omega/T}}\)e^{-\omega/T}+\frac{1}{2}\frac{e^{\omega/T}-2}{e^{\omega/T}-1}\Gamma^{\rm P,C}_a+\Gamma^{\rm P,ee}_a \\
&=& 
\frac{1}{2}\frac{g_{ae}^2 \omega^2}{e^2 m_e^2}\frac{k(\omega)}{e^{\omega/T}-1} +\frac{1}{2}\frac{e^{\omega/T}-2}{e^{\omega/T}-1}\Gamma^{\rm P,C}_a+\Gamma^{\rm P,ee}_a 
\label{eq:ABCfromOPA}
\eea

The first term, proportional to the photon opacity, includes the contribution from ff, fb and bb processes and 
we have obtained it essentially from the proportionality factor of case I, eq. ~\eqref{eq:gammas}.  
The same term includes also a part of the Compton contribution $\Gamma^{\rm P,C}_a$ but not all   
because is not ruled by~\eqref{eq:gammas} and because $\Gamma^{\rm A,C}_\gamma$ contributes to the opacity without the stimulation factor $(1-e^{\omega/T})$. The second term is there to complete the Compton contribution. 
Compton is only relevant for $\omega/T\gtrsim 5$ so the second term is one half of the whole Compton production rate of axions.  
The third contribution is the axion emission in e-e bremsstrahlung, which has to be added apart.

\section{Solar axion Flux}

The differential axion flux from the Sun is obtained by integrating the axion emission rate $\Gamma^{\rm P}_a$  times the phase space density over the volume of the Sun
\be
\frac{d\Phi_a}{d\omega}=\frac{1}{4\pi R_{\rm Earth}^2} \int_{\rm Sun} dV \frac{4\pi \omega^2}{(2 \pi)^3} 
\Gamma^{\rm P}_a(\omega)
\ee
$\Gamma^{\rm P}_a(\omega)$  depends on the position on the Sun through its dependence on the local characteristics 
of the plasma: $T,\rho,X_Z$ for temperature, mass-density and mass-fraction of the chemical element $Z$.  
The latter are taken to depend only on the radial position inside the Sun and given by a solar model. 
In this paper we have used the solar model AGSS09~\cite{Serenelli:2009yc}, available at~\cite{AGSS09}, which was calculated after the last revision of the solar atomic abundances by Asplund et al.~\cite{Asplund:2009fu}. 
The solar model includes element diffusion  (the chemical composition depends on the solar radius) and utilises the radiative opacities of the Opacity Project (OP)~\cite{OP}. In the next subsections we compute the solar emission of axions with OP opacities and compare it with those obtained with two different opacity codes: LEDCOP and OPAS.  

\subsection{Opacity Project}

The Opacity Project is an international collaboration aiming at calculating extensive atomic data required to estimate radiative opacities based on the Hummer-Mihalas formalism~\cite{Hummer1988} for the equation of state of stellar plasmas. It is intended to provide accurate opacities for relatively low densities and temperatures like those present in stellar envelopes~\cite{Seaton1994}. Under these conditions, it is reasonable to assume that plasmas do not affect the structure of atoms although they do impact the probability of finding a given atomic configuration. The radiative opacity of the plasma is thus calculated by computing free-free and bound-free cross-sections as if they would happen in vacuum, multiplied by the probability (which depends on the density, composition and temperature) of finding the correct initial states. The code includes plasma corrections only in bound-bound transitions through their line broadening and in Compton scattering (which are largely irrelevant for our purposes). 
This simplified picture allows the OP to make monocromatic opacities available in a very flexible way by splitting the absorption coefficient in a sum over atomic nuclei  
\be
k(\omega, T,\rho,\{X_Z\},n_e)\equiv \sum_Z n_Z \sigma^{\rm OP}_Z(\omega,T,n_e) (1-e^{\omega/T})
\ee
where $n_Z$ is the density of atoms with atomic number $Z$.  
Here $\sigma^{\rm OP}_Z(\omega;T,n_e)$ is an effective absorption cross section that includes ff, bf, bb thermal averaged cross sections summed over possible ionization stages and excited states (all these quantities depend on $\omega,T,n_e$ but not on $n_Z$) and scattering. The contribution of scattering is not proportional to $n_Z$ but to $n_e$, but both quantities are related through the averaged ionization fractions $I_Z$ (which again, within the approximations of the OP, depend on $T,n_e$ but not on $n_Z$)
\be
n_e = \sum_Z I_Z  n_Z , 
\ee
so that each $\sigma^{\rm OP}_Z$ includes the contribution of electrons that in a non-ionized plasma would be in atoms of atomic number $Z$. 

Since their first release, the OP has continued to update their opacities including new reactions and atomic proceses. Inclusion of inner-shell processes brought a remarkable agreement with OPAL opacities~\cite{Badnell2003} even in high-temperature and density regions where the OP calculations where not originally intended to be accurate. After inclusion of inner-shell processes for all cosmically abundant elements the agreement is remarkable for frequency-averaged $k(\omega)$~\cite{Badnell2004}. We will attempt an explanation for this fact below.  

\begin{figure}[htbp]
\begin{center}
\includegraphics[width=7cm]{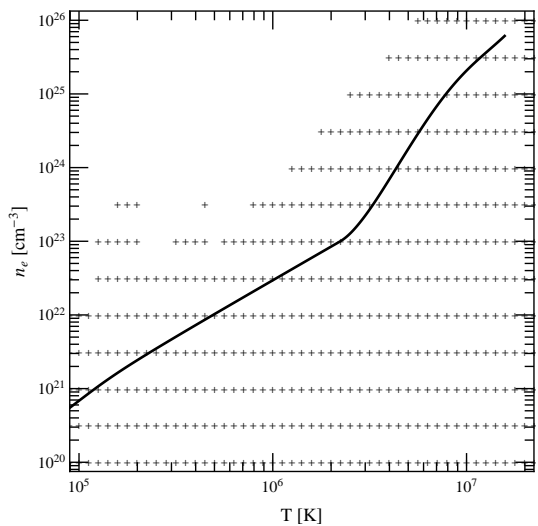}
\caption{Solar model (black line) and points of temperature and electron density $n_e$ where monochromatic opacities from the OP are available. }%\small Reactions responsible for the }
\label{fig:mesh}
\end{center}
\end{figure}

The OP provides tables of $\sigma_Z(\omega)$ for the 17 cosmologically abundant elements (H, He, C, N, O, Ne, Na, Mg, Al, Si, S, Ar, Ca, Cr, Mn, Fe and Ni) in a logarithmic grid of temperatures  $T\in(10^{3.5},10^{8})$ K and a maximum range of electron densities $n_e\in(10^{3.5},10^{29})$cm$^{-3}$ (the range depends on the temperature)~\cite{Seaton:2004uz}. 
The available grid covers nicely our solar model, see figure~\ref{fig:mesh}, but the coverage is a bit too short in density\footnote{A finer mesh with twice as many points exists but it is not available in the public release. } (only two points per decade). In order to integrate the solar axion emission we performed an interpolation of the OP data in logarithmic units to compute $k(\omega)$ in a grid of values corresponding to the solar model.  

\begin{figure}[tbp]
\begin{center}
\includegraphics[width=8cm]{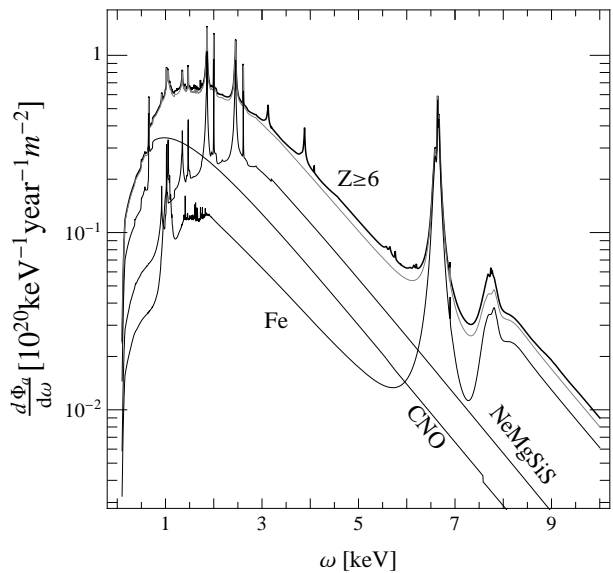}
\caption{Solar axion flux ($g_{ae}=10^{-13}$) from reactions that involve all cosmologically abundant metals ($Z\geq 6$, upper black) and three groups that account for most of the flux: CNO, NeMgSiS and Fe. The sum of the latter adds up to the think gray line which tracks to a good approximation the full result.  }%\small Reactions responsible for the }
\label{fig:metals}
\end{center}
\end{figure}

\begin{figure}[htbp]
\begin{center}
\includegraphics[width=7cm]{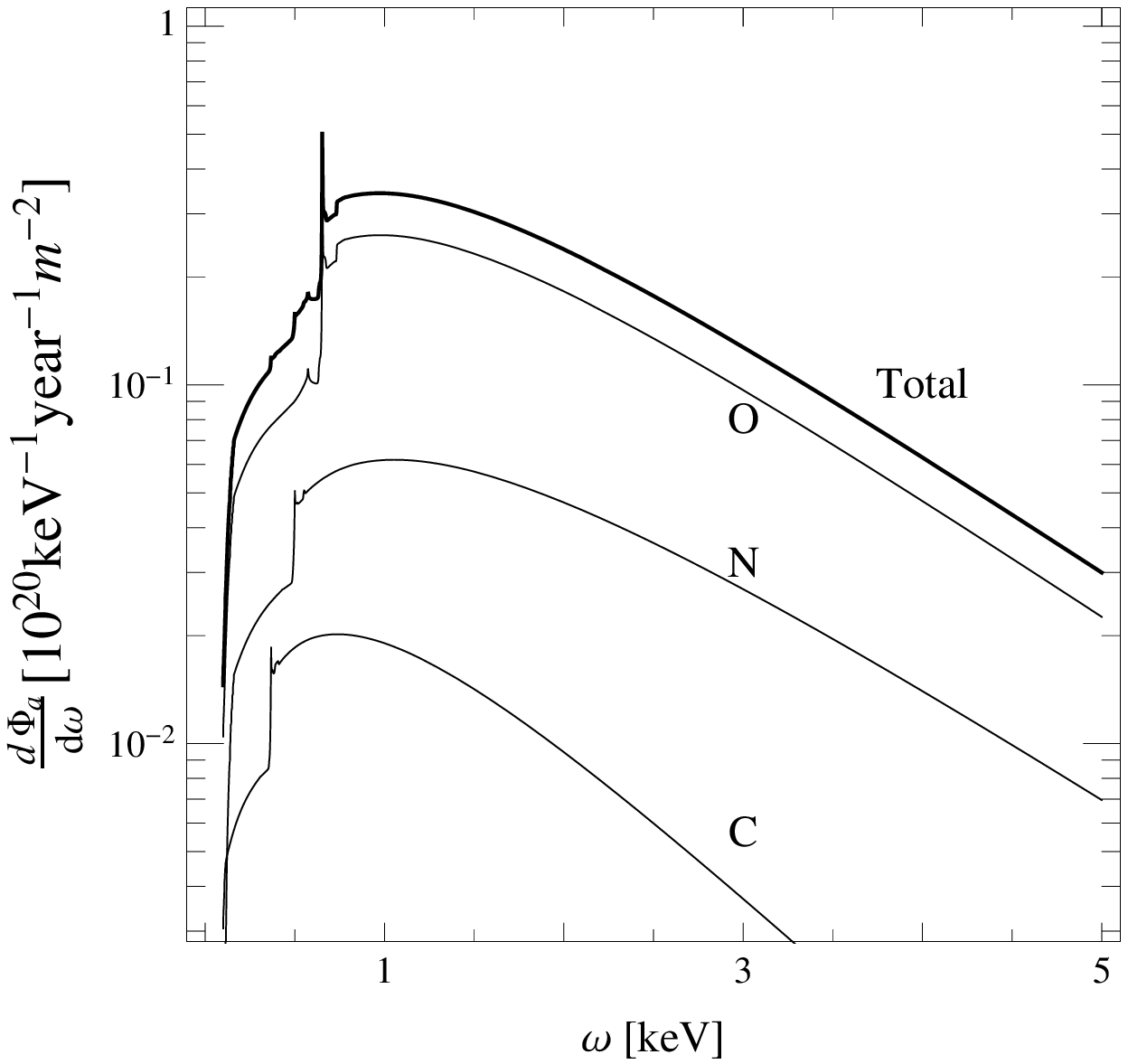}
\includegraphics[width=7cm]{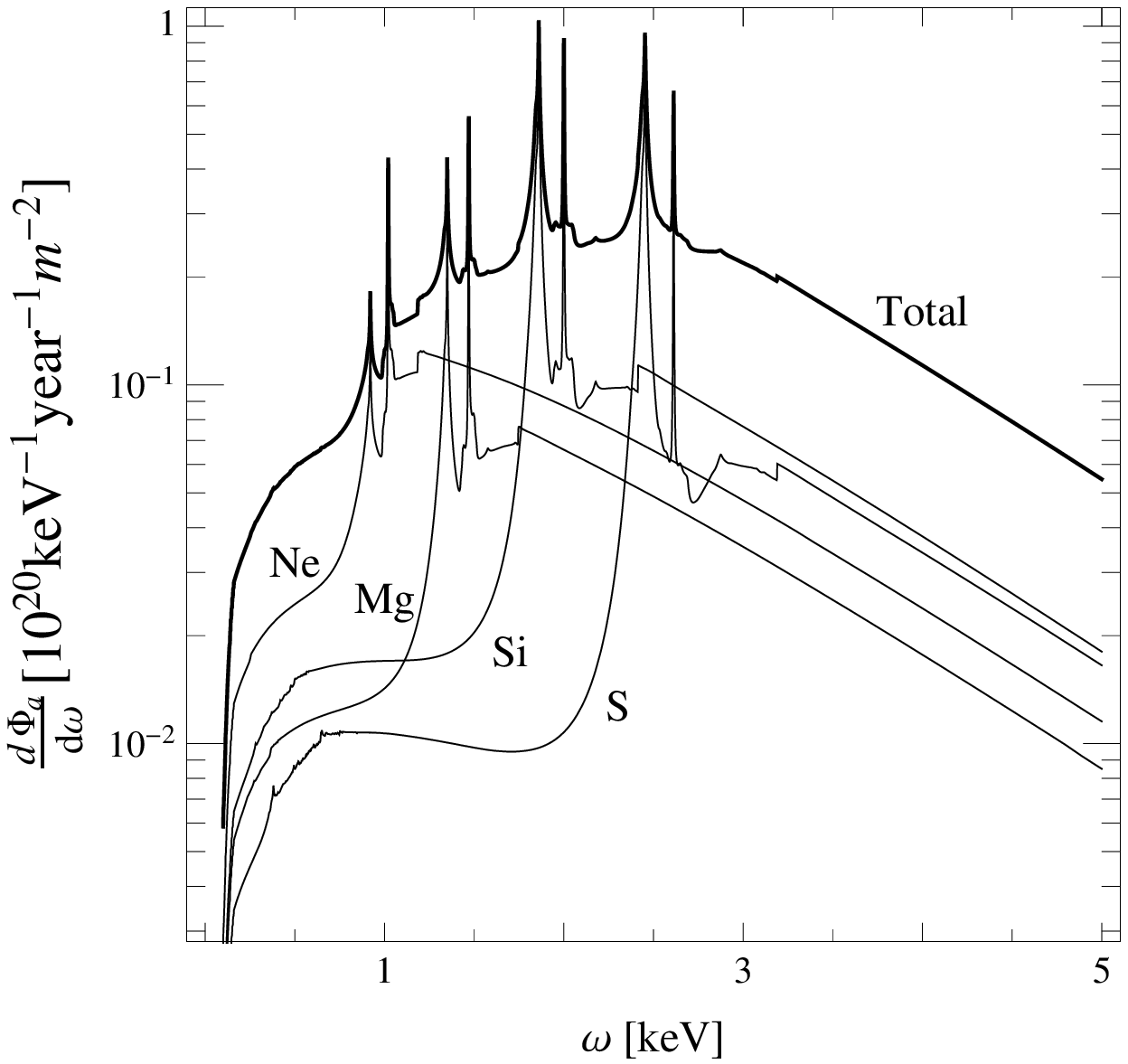}
\caption{Solar axion flux ($g_{ae}=10^{-13}$) from reactions that involve CNO (left) and NeMgSiS (right)}%\small Reactions responsible for the }
\label{fig:metals2}
\end{center}
\end{figure}

OP opacities are given for each chemical element independently so we can split the solar axion flux 
in components in which only one chemical element was involved. The contribution from metals $Z\geq 6$ is the novel part to discuss here. In figure~\ref{fig:metals} we have plotted the axion flux at Earth 
due to reactions involving nuclei with $Z\geq 6$ (black line) and split it into three groups of elements 
with important contributions: CNO, (blue line), NeMgSiS (red line) and Fe (magenta line). These elements add up almost to the full result. Other metals have very small abundances and contribute very little. 
Details from the CNO and NeMgSiS with their individual contributions are shown in figure~\ref{fig:metals2}. 

%The flux is dominated by the emission of axions from the solar center, up to a radius of 20-30\% of the solar radius. 
%This is because the emission of axions is a relatively strong function of the atomic densities and the temperature. 

If we focus on the contribution of one element we can observe spectral features of the different 
processes contributing to the axion flux. 
At the highest energies the dominating feature is the electron recombination into the K-shell (free-bound). 
This emits axions with a minimum energy given by the K-shell energy and thus has a sharp break 
at this energy. 
At slightly lower energies one can see very prominently the Ly-$\alpha$ emission (bound-bound) where an electron 
in the 2p orbital de-excites into the 1s. Between this prominent peak and the K-shell threshold, the rest of the 
Lyman series can be found (de-excitations from more energetic orbitals) but can rarely be seen due to merging of the lines. Below Ly-$\alpha$ lines the spectrum decreases and gets typically dominated by electron-recombination into the L-shell whose threshold cannot be seen in any of the examples shown. Just below the threshold, one could in principle see the Balmer series but these two last features can be seen only in the case of Fe (in the range 1-2 keV). 
The contribution from bremsstrahlung (free-free) would be featureless and could not be easily identified. 
However, the prominency of features tells us that it is subdominant with respect to the free-bound and bound-bound transitions commented above.  
Thus we can conclude that most of the axion flux from reactions involving metals is due to free-bound and bound-bound processes. 
 
\begin{figure}[tbp]
\begin{center}
\includegraphics[width=9cm]{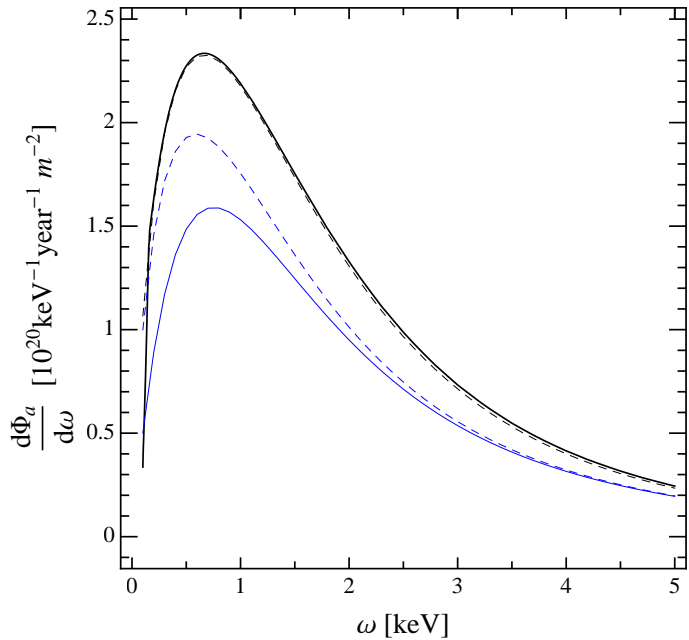}
\caption{Solar axion flux ($g_{ae}=10^{-13}$) from reactions involving H and He ions together with the Compton process. Calculation from the OP opacity (black solid), from eq. ~\eqref{eq:axioBremsstrahlung} including screening (blue solid) and excluding screening (blue dashed) and finally without screening and the vacuum Gaunt factor of Karzas and Latter (Black dashed). }
\label{fig:ZeCompton}
\end{center}
\end{figure}

Let us now turn into the contribution from H and He, the most abundant elements in the Sun. 
The energy levels of H and He are much below the X-ray energies that dominate the flux of 
solar axions and thus we do not expect any features in the resulting flux. The result should therefore 
be exclusively given by the free-free contribution (axio-bremsstrahlung) and Compton (remember that 
we cannot separate them, they are given together in the OP data).  Since we have good 
analytical formulas for the axion emission, \eqref{eq:axioBremsstrahlung} and \eqref{eq:axioCompton}, we can compare the results obtained by using OP opacities to a direct integration of the rates. 
The spectra are shown in figure~\ref{fig:ZeCompton}. The black line is the flux obtained from the OP data while the blue thick line is the one from expressions \eqref{eq:axioBremsstrahlung} and \eqref{eq:axioCompton}. 
We see that the latter is considerably reduced with respect to the former. 
Naively, one could think that this is due to the fact that \eqref{eq:axioBremsstrahlung} and \eqref{eq:axioCompton} include Debye-screening while the OP opacities do not. 
If we use \eqref{eq:axioBremsstrahlung} and \eqref{eq:axioCompton} by removing screening (i.e. we use $k_s=0$) the resulting flux is still too small (blue dashed line). 
The solution to this conundrum is that the OP free-free opacities are computed using the classical Kramers formula times the full vacuum Gaunt factor, as computed from Karzas and Latter~\cite{Karzas:1961}. This 
Gaunt factor includes corrections from the fact that electron-wavefunctions are distorted by the nuclei they are scattering from. Electron wave-functions are larger closer to the nucleus and this tends to enhance the final cross section. Equation~\eqref{eq:axioBremsstrahlung} is computed in the Born approximation where electrons are taken to be plane waves and therefore does not feature this enhancement. 
This conclusion is based upon a third calculation, shown as a dashed red line in figure~~\ref{fig:ZeCompton} 
where we have used our expression~\eqref{eq:axioBremsstrahlung} modified to include the vacuum Gaunt factor of Karzas and Latter. This result fits perfectly with the OP calculation. 
This is one of the undesirable features of using OP opacities, designed for low dense plasmas, for the plasma of the deep solar interior. 
The Gaunt factor used by the OP is not justified in the deep Sun because Coulomb wave-functions have modifications over typical distances given by the size of the electronic orbitals. In the case of H and He, the sizes are much bigger than the average inter-electron distance and thus the vacuum treatment is not justified at all. 
In this paper we chose to represent the free-free contribution by the Born approximation with Debye screeening formula~\eqref{eq:axioBremsstrahlung}, which is more adequate to the solar interior than the OP. 
This is a reasonable choice and, moreover, it is supported by the calculations based on Los Alamos and OPAS opacities which we present below. 

Note that plasma effects affect much less the atomic transitions in the relevant metals ($Z\geq 6$) simply because 
the corresponding orbitals are much smaller. The previous study of the free-free contributions shows that 
even if there are differences between treating atoms in vacuum or in a plasma, or including Coulomb-wavefunctions for the scattering states or plane waves, the differences are relatively small for H and He (30\% at the very most in figure~\ref{fig:ZeCompton}). 
We expect them to be even smaller for the processes involving metals.  

The best calculation of the ABC flux of solar axions from the OP opacities is therefore obtained by 
adding up the free-free contribution from~\eqref{eq:axioBremsstrahlung} and e-e bremsstrahlung contribution from~\eqref{eq:axioBremsstrahlungee}, Compton from~\eqref{eq:axioCompton}  and the contribution from metals from the OP opacity shown in figure~\ref{fig:metals}.  
The resulting spectrum is shown in figure~\ref{fig:flux} together with the different components mentioned above.  

%\subsection{Comparison with other opacity codes}

As a cross check of our calculations we have used different opacity codes to compute the flux of ABC axions. 
In general we find very good agreement on smoothed spectra, but (somewhat irrelevant) discrepancies are evident in the position and intensity of the emission lines. This is a well known fact that derives from the different opacity codes, which use different methods to compute the atomic data (energy levels, dipole matrix elements, occupation probabilities), see e.g.~\cite{TurckChieze:2011ht}.  

Small discrepancies are also to be expected in the low energy part of the spectrum $\omega <2$ keV where 
plasma screening plays a role in the free-free contribution around the solar center. 
This region is not relevant for radiative opacities because the quantity relevant for stellar evolution is the 
frequency-averaged Rosseland opacity, 
\be
\frac{1}{\kappa_{\rm R}}=\frac{15}{4\pi^4} \int_0^\infty \frac{\omega^4 d \omega\, e^{\omega/T}}{\kappa(\omega) (e^{\omega/T}-1)^2}
\ee
which favours frequency regions of small $\kappa(\omega)$ (and below 2 keV the opacity is huge) and it is relatively insensitive to the region $\omega/T <2$.  Thus opacity codes do not need to put much effort in this region. Notwithstanding the above, it turns out that even here different codes agree reasonably well. 

\subsection{Los Alamos opacities (LEDCOP)}

Elemental opacities computed with the Los Alamos LEDCOP Opacity Code~\cite{LEDCOPref} can be combined by the TOPS code~\cite{TOPS} and are available online in~\cite{LEDCOP}. 
In LEDCOP, each ion stage is considered in detail in a similar way to the OP code, but here interactions with the plasma are included as perturbations. Energy levels and radial dipole matrix elements are computed in LS or in intermediate coupling and fitted to simpler formulas which are finally used in the code. 
The equation of state is calculated iteratively solving the Saha equation and the bound Rydberg sequences are cut off by plasma corrections. The free-free absorption is calculated from the Kramers cross section
formula and relativistic free-free Gaunt factors~\cite{Nakagawa1987}, which amount to very small corrections in our region of interest.  
%The Los Alamos Opacity Library has been offering astrophysical opacities 
  
\begin{figure}[htbp]
\begin{center}
\includegraphics[width=12cm]{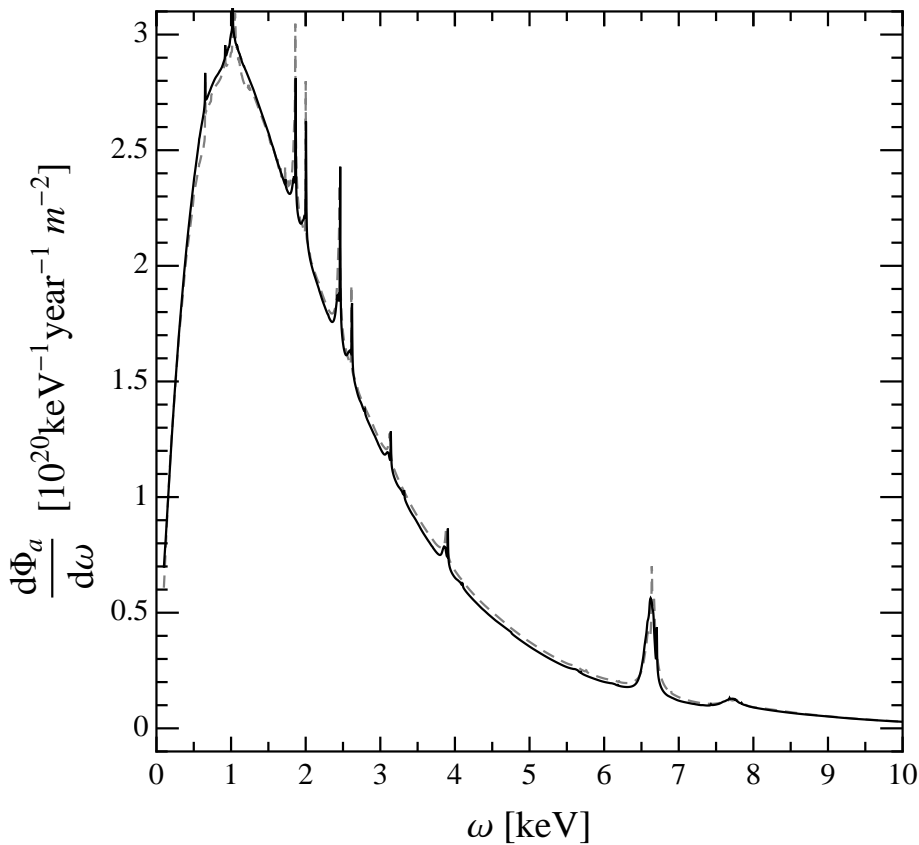}
\caption{Solar ABC flux computed with LEDCOP opacities (black) compared with the OP (gray).}%\small Reactions responsible for the }
\label{fig:LAOL}
\end{center}
\end{figure}

LEDCOP opacities are offered in a sparse array of temperature (only 11 temperature points from 0.1 to 1.25 keV, our region of interest) for arbitrary density and include all light elements $Z<31$. 
In order to integrate over a solar model we have created monochromatic opacities for intermediate temperatures by interpolating the temperature grid.  
Integration over the solar model of the axion emission formula~\eqref{eq:ABCfromOPA} gives the solar axion ABC flux shown in figure~\ref{fig:LAOL} as a solid line. For comparison we have also displayed the OP-corrected flux obtained in the previous section.   
The agreement is really remarkable except close to the lines caused by bound-bound processes. However, both the OP and LEDCOP groups warn that their opacities are not spectroscopically resolved, so some difference was to be expected. For the current work this discrepancy is not relevant, as we are very far from being able to to axion spectroscopy and bound-bound processes contribute little to the total flux.   

As a matter of fact, we find the agreement too good. We have shown before that free-free processes did not agree between the OP and our direct calculation using Debye-screened inverse-bremsstrahlung in the Born approximation. LEDCOP uses essentially the same Gaunt factors that the OP (relativistic corrections are very small in the solar core and electron degeneracy corrections are also small) and we should have expected to show a bit more prominently (around 0.7 units of figure~\ref{fig:LAOL}, i.e. $\sim 25\%$ at the peak at $\omega\sim$ keV). 
This putative discrepancy can be explained by a decrease of the free-bound processes at low energies 
but could also be due to an unaccurate number of points in our integration over the solar interior or our interpolation of the opacities.

\subsection{OPAS}

We have performed a further cross-check by using the opacities calculated with the OPAS\footnote{I am very grateful to C.~Blancard for sharing the OPAS data and clarifying some issues concerning the free-free contribution.} code~\cite{OPAS}.  
OPAS uses the average-atom model SCAALP~\cite{SCAALP} to compute self-consistently energy levels 
and average ocupation numbers of detailed atomic configurations. Polarization and correlation effects of the continuum electrons are taken into account. 
OPAS includes free-free, bound-free and bound-bound opacities as well as scattering. 

\begin{figure}[htbp]
\begin{center}
\includegraphics[width=12cm]{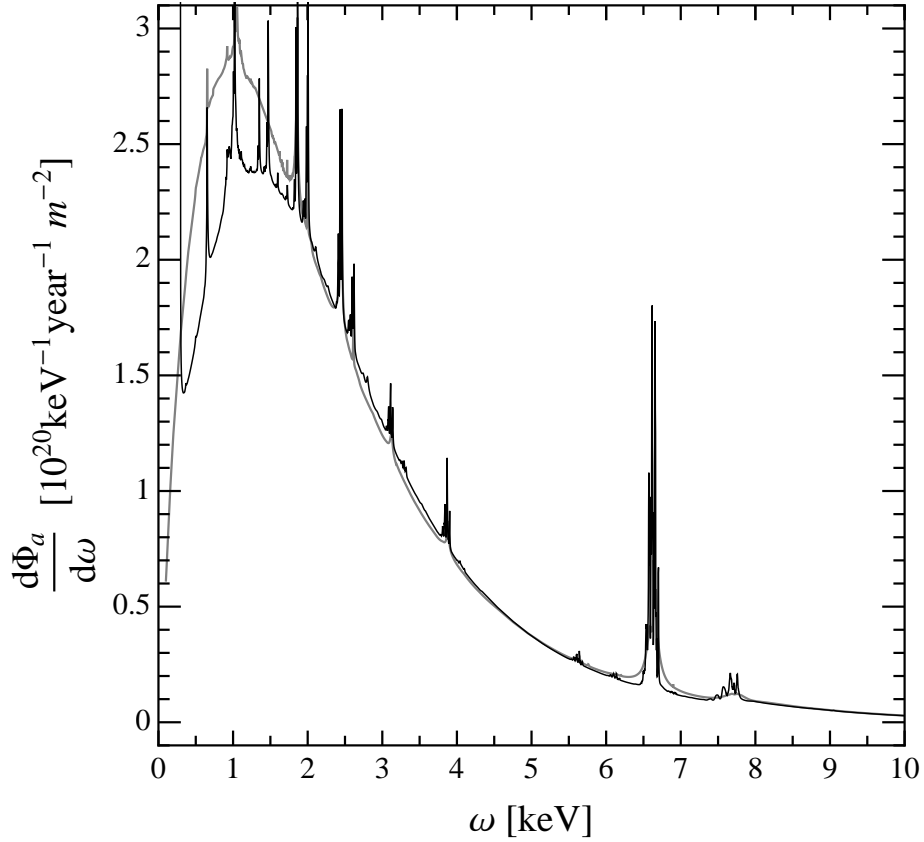}
\caption{Solar ABC flux computed with OPAS opacities (black) compared with the OP (gray).}%\small Reactions responsible for the }
\label{fig:OPAS}
\end{center}
\end{figure}

The OPAS team performed recently a calculation of monochromatic opacities following the temperature and density dependence of the Sun~\cite{OPAS}, using the recently revised chemical abundances of the Sun~\cite{Asplund:2009fu}. Using these data we can compute the OPAS flux of ABC axions by integrating eq.~\eqref{eq:ABCfromOPA} over the solar model. The result is shown in figure~\ref{fig:OPAS}.  
Again, the results compare extremely well with the OP calculation (shown as a gray line). 
The contribution from bound-bound processes is more prominent that in the OP or the LEDCOP and is better resolved,  
but the differences are not significant when smoothed over frequencies. 

The most remarkable feature of the OPAS result is the suppression of the emission below $\sim 2$ keV 
and the abrupt increase at 300 eV. The first effect is due to the free-free contribution to the opacity, which at low 
energies is interpolated to match the low-frequency regime of a plasma, given by the Drude conductivity~\cite{QMD}. 
The second is due to the plasma-frequency. Photons with a frequency smaller than the plasma frequency ($\omega_\text{P}\sim 300$ eV in the solar center) cannot propagate freely (just as massive particles with energies below their mass), which OPAS interprets as a contribution to the absorption coefficient $k={\rm Im} \sqrt{\omega^2-\omega_\text{P}^2}$. Both these effects have no counterpart for axions and are undesirable for our purposes. Thus, the region below 2 keV cannot be used in earnest to estimate the ABC flux.  
 
%%%%%%%%%%%%%%%%%%%%%%%%%%%%%%%%%%%%%%%%%%%%%%%%%%%%%
\section{Beyond axions}
%%%%%%%%%%%%%%%%%%%%%%%%%%%%%%%%%%%%%%%%%%%%%%%%%%%%%

The ideas presented in this work have direct implications for other novel weakly interacting slim particles (WISPs). 
The obvious example is that of axion-like particles (ALP). 
Symmetries such as the U$_{\rm PQ}$(1) and their corresponding pseudo-Goldstones can be easily implemented by hand or they might appear as unwanted (and extremely predictive) features of field theoretic extensions of the standard model.  
%Reference models such as the minimal KSVZ or the DFSZ have served to guide experimental searches and constraint different cosmological scenarios involving axions. Other axion-like particles have been proposed as a result of spontaneous breaking of  
Moreover, in recent years it is becoming increasingly clear that these symmetries are generic predictions of the most ambitious high-energy completion of the standard model, string theory. What is more, a typical string compactification can cast a vast number of axion-like particles (ALPs)--- typically O(100) --- out of which one can be the QCD axion. 
From the phenomenological point of view, this proliferation of ALPs is very welcome. 
Axions and/or axion-like particles can provide satisfactory explanations to many observations: absence of CP violation in strong interactions, dark matter, dark radiation, anomalous-transparency of the universe to gamma-rays or white-dwarf cooling but the explanations often require different parameters, such as decay constants or masses. 
Thus, our bottom-up model building is somehow supported by the most speculative top-down reasoning (perhaps unsurprisingly). 

From the phenomenological point of view, these ALPs behave much like axions but their mass is not related a-priori to their decay constant (which set's natural values for the couplings).
If these particles have an ALP-electron coupling, our predictions apply mutatis mutandis to them, given that their mass is below the keV.  

New vector bosons can be also constrained with these results. Hoffmann showed that if a novel photon-like species, a hidden photon (HP), has a magnetic coupling to electrons
\be
{\mathcal L}_{M\gamma'} = \frac{g_{M\gamma'}}{4 m_e} F'_{\mu\nu }\bar \psi_e \sigma^{\mu\nu} \psi_e
\ee
where $F'_{\mu\nu}=\partial_\mu A'_\nu-\partial_\nu A'_\mu$ is the field strength of the hidden photon field $A'_\mu$, 
then the spin-averaged cross sections for hidden photon emissions are proportional to those of axion emission~\cite{Hoffmann:1987et}. 
The flux of magnetically coupled hidden photons is given by 
\be
\frac{d\Phi_{M\gamma'}}{d\omega}=2 \(\frac{g_{M\gamma'}}{g_{ae}}\)^2\frac{d\Phi_{a}}{d\omega} . 
\ee

If the coupling is of electric type, i.e. 
\be
{\mathcal L}_{E\gamma'} = g_{E\gamma'} A'_\mu\bar \psi_e \gamma^{\mu} \psi_e
\ee
the situation is very delicate and depends much on the hidden photon mass and the coupling of the hidden photon to the rest of standard model particles. Recent papers also showed that in some cases the emission of longitudinally polarized modes can dominate~\cite{An:2013yfc,Redondo:2013lna}. It is definitely beyond the scope of this paper to explore the different possibilities. 

\section{Acknowledgements}

The author is very thankful to Christophe Blancard and the OPAS team for making available their opacities for this paper,  to Georg~Raffelt for multiple discussions on its topic and a careful reading and to Igor Irastorza for the stimulation required to jump on this project.  
It is a pleasure to acknowledge support by the Alexander von Humboldt Foundation and 
 by the European Union through the Initial Training Network ``Invisibles''.

\end{document}